%% file: main.tex
\documentclass[10pt,conference]{IEEEtran}
\usepackage[T1]{fontenc}
\usepackage{cite}
\usepackage{xurl}        
\usepackage{amsmath,amssymb}
\usepackage{graphicx}
\usepackage{booktabs}
\usepackage[hidelinks]{hyperref}

\usepackage[most]{tcolorbox}

\newtcolorbox{rqtakeaway}{
    enhanced,
    breakable,
    colback=white,
    colframe=black!40,
    boxrule=0.5pt,
    arc=1mm,
    left=6pt,
    right=6pt,
    top=5pt,
    bottom=5pt,
    before skip=6pt,
    after skip=6pt
}

\title{Beyond Leaderboard Ranks: Selecting Software Engineering Agents Under Budget Constraints}
\begin{document}
\bstctlcite{IEEEexample:BSTcontrol} 
\author{
  \IEEEauthorblockN{Srinath Srinivasan}
  \IEEEauthorblockA{Dept. of Computer Science\\
  North Carolina State University\\
  Raleigh, North Carolina\\
  Email: ssrini27@ncsu.edu}
  \and
  \IEEEauthorblockN{Tim Menzies}
  \IEEEauthorblockA{Dept. of Computer Science\\
  North Carolina State University\\
  Raleigh, North Carolina\\
  Email: timm@ieee.org}
}
\maketitle

\begin{abstract}
Teams that adopt LLM agents for issue resolution must decide which agent
to pay for, and how much to spend finding out.
Leaderboard ranks give weak evidence for this decision.
Accuracy differences may be statistically unresolved,
reported costs may be inconsistent, and evaluating
candidate agents can itself be expensive.
We investigate budget-constrained agent selection using
58,197 run--task records extracted from the Holistic
Agent Leaderboard (HAL), focusing on SWE-bench and
six additional benchmarks.

Across seven benchmarks, 27 of 35 steps along the
accuracy--cost Pareto frontiers are statistically
unresolved.
On SWE-bench, a run costing \$34.16 trails the
\$170.78 leader by 14.6 percentage points, but
the available tasks cannot statistically distinguish
their accuracy.
Reported costs also differ from standardized
token-based estimates by up to 11.7$\times$,
while selective task evaluation can bias estimates
of full-benchmark execution cost.

We therefore compare evaluation strategies under
explicit evaluation budgets and deployment-cost caps.
Under advance knowledge of task costs, successive
halving reaches our uncertainty-aware selection target
on SWE-bench using 6.6\% of exhaustive evaluation
spend, achieving the lowest cost on six of seven
benchmarks and tying on the seventh.
Its advantage narrows when task costs are unknown
before execution or exact-best selection is required.

Our findings motivate decision-oriented agent
evaluation that accounts for accuracy uncertainty,
validates execution costs, and compares complex
selection methods against simple adaptive allocation.

\end{abstract}

\section{Introduction}

Software teams now use autonomous agents to resolve issues in their
repositories~\cite{agentless, sweagent, wang2025openhands}.
Each agent combines a language model with a \emph{scaffold} that navigates
the code, edits files, and runs tests.
A team that adopts such an agent faces a maintenance budget question:
which agent gives the most resolved issues per dollar on its backlog,
and how much should the team spend to find that out?
Two budgets apply.
The \emph{evaluation} budget pays for trial runs of candidate agents.
The \emph{deployment} budget pays for the chosen agent on future issues.
Both matter, because one multi-turn repair attempt can consume many
thousands of tokens, and small per-issue cost differences grow large
over a backlog of hundreds of issues.

Existing evaluations primarily compare agents through leaderboard
rankings.
Prior work has advocated cost-controlled agent benchmarking~\cite{agentsmatter}
and proposed task-subset sampling methods that preserve leaderboard
rankings while reducing evaluation effort~\cite{ndzomga}.
However, reproducing a ranking does not necessarily identify the
best agent for a given budget.
Observed accuracy differences may be statistically unresolved, while
reported or sampled execution costs may not reflect the expense
of running an agent across the full benchmark.
This raises a practical question:
\emph{How can we select a competitive, affordable software engineering
agent when evaluation itself is constrained by a dollar budget?}

\begin{figure*}[t]
\centering
\includegraphics[width=0.8\textwidth]{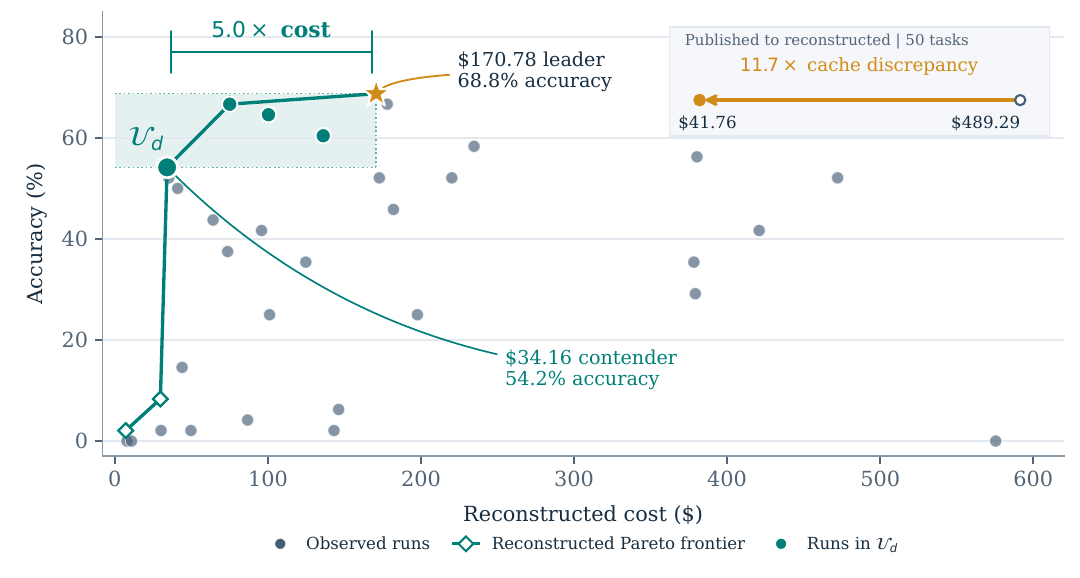}
\caption{Reconstructed API cost and accuracy of 34
SWE-bench Verified Mini runs on 48 shared tasks.
The teal line traces the observed accuracy--cost
Pareto frontier. At deployment cap $d=\$170.78$,
the affordable unresolved set $\mathcal{U}_d$
contains the leader and four cheaper runs whose
paired accuracy differences from the leader
remain unresolved after BH correction ($q=0.05$).
The \$34.16 contender costs 5.0$\times$ less
than the leader. Shading indicates the observed
accuracy range. The inset shows a separate 50-task
run whose published cost (\$489.29) is
11.7$\times$ its reconstructed cost (\$41.76)
due to prompt-cache accounting discrepancies.}
\label{fig:swe-tradeoffs}
\end{figure*}

We investigate this question using execution traces from the
Holistic Agent Leaderboard (HAL)~\cite{hal}, focusing on
SWE-bench Verified Mini~\cite{swebench,sweverified}.
Across 48 shared tasks, the highest-scoring run costs
\$170.78, while a \$34.16 contender trails by
14.6 percentage points.
Yet their paired accuracy difference is statistically
unresolved, making it unclear whether the fivefold
cost premium is justified.
Figure~\ref{fig:swe-tradeoffs} illustrates this trade-off
and shows how inconsistent cost accounting can further
distort agent comparisons.

Motivated by this observation, we study agent selection as a
\emph{budget-constrained decision problem} rather than a
leaderboard-ranking problem.

For repository-maintenance agents, this decision
is shaped by the software repair workflow itself.
Agents may repeatedly inspect source files, modify
patches, execute tests, and diagnose failures,
making execution cost dependent on the task and
the agent's scaffold.
Consequently, selecting an agent requires comparing
complete model-scaffold configurations under
uncertain accuracy and execution costs, rather than
ranking language models in isolation.

We address three research questions:

\begin{itemize}

\item \textbf{RQ1: Which accuracy improvements among more expensive
agents can benchmark evaluations statistically resolve?}
Across seven benchmarks, 27 of 35 steps along the reconstructed
accuracy--cost Pareto frontiers remain statistically unresolved.
This limits what leaderboard rankings can establish about
the benefits of paying for higher-ranked agents.

\item \textbf{RQ2: How reliable are reported and sampled estimates
of agent execution cost?}
Leaderboard costs differ from standardized token-based estimates
by as much as 11.7$\times$, while selectively evaluating
medium-difficulty tasks can underestimate full-benchmark
execution costs by 20.4\% on SWE-bench.
Both effects complicate cost-aware agent selection.

\item \textbf{RQ3: Which evaluation strategies efficiently identify
affordable, competitive agents under a dollar budget?}
We compare simple allocation strategies with more complex
acquisition and stopping methods.
Under advance knowledge of individual task costs,
successive halving reaches our uncertainty-aware selection target
using 6.6\% of exhaustive SWE-bench evaluation spend.
It achieves the lowest spend on six of seven benchmarks and
ties on the seventh, although its advantage narrows when task
costs are unknown before execution.

\end{itemize}

To answer these questions, we extract 58,197 run--task records
from HAL, reconstruct costs using standardized token accounting,
analyze paired accuracy differences, and replay evaluation
strategies under varying dollar budgets and cost-visibility
assumptions.
Rather than introducing a new optimization algorithm,
our contribution is an empirical assessment of the evidence
and evaluation strategies needed for budget-constrained
agent selection.

\smallskip\noindent
\textbf{Scope.}
SWE-bench is our primary case.
It is the only repository-level maintenance benchmark in HAL with
complete binary outcomes and recoverable token traces.
To check whether our findings depend on this one benchmark, we repeat
the analysis on three programming benchmarks (CORE-Bench~\cite{siegel2024core},
SciCode~\cite{tian2024scicode}, USACO~\cite{shi2024can}) and summarize three
general tool-use benchmarks (GAIA~\cite{mialon2024gaia},
TAU-bench~\cite{yao2025taubench}, AssistantBench~\cite{yoran2024assistantbench}).
Our software maintenance conclusions rest on SWE-bench alone.

\section{Background and Related Work}

\subsection{Software-Maintenance Benchmarks}

Repository-level benchmarks evaluate software engineering agents
by testing whether generated patches resolve real defects~\cite{swebench}.
Repair success depends not only on the underlying language model
but also on the agent's scaffold and tool-use strategy.
SWE-agent demonstrates the importance of agent--computer
interfaces~\cite{sweagent}, while Agentless shows that a fixed
localization-and-repair pipeline can remain competitive without
autonomous tool loops~\cite{agentless}.
SWE-bench Verified contains 500 human-screened repository
issues~\cite{sweverified}, and the Holistic Agent Leaderboard
(HAL)~\cite{hal} uses a 50-task Verified Mini subset to reduce
evaluation costs.

These benchmarks support comparisons between agent designs,
but their rankings alone do not establish whether additional
accuracy justifies higher execution costs.
We study this selection problem using the 48 SWE-bench tasks
shared by all complete runs in our retained HAL data
(Section~\ref{sec:data}).

\subsection{Effort-Aware Evaluation in Software Engineering}

Software engineering has a long history of evaluation that weighs
benefit against the effort needed to obtain it.
Effort-aware defect prediction ranks modules by predicted defects per
line inspected, not by raw defect probability~\cite{mende2010effort,
kamei2012large}.
Under this measure, simple unsupervised predictors can match or beat
complex supervised ones~\cite{yang2016effort, fu2017revisiting}.
We apply the same view to agent selection.
The ``effort'' is the dollar cost of agent execution, and the
``benefit'' is the number of resolved issues.
As in effort-aware defect prediction, we test whether simple allocation
rules match more complex ones once cost enters the evaluation.

Prior work also questions whether leaderboard ranks give enough
evidence for model selection~\cite{card2020little, miller2024adding,
dad, dior}.

\subsection{Efficient Benchmark Evaluation}

Existing approaches reduce evaluation effort by selecting
informative benchmark examples.
Anchor Points identifies representative subsets that preserve
leaderboard rankings~\cite{anchorpoints, polo2024tinybenchmarks}, while Active Testing
adaptively selects examples and corrects for sampling
bias~\cite{activetesting}.
Closer to our setting, Ndzomga et al.~\cite{ndzomga} study task-subset selection
across agent scaffolds and models, including mid-range filtering
that prioritizes tasks with intermediate historical pass
rates.

Other approaches predict unobserved benchmark outcomes.
Li et al.~\cite{aea} use learned dependencies between examples to guide
evaluation acquisition, while Zhang et al.~\cite{misses} find
that simple random sampling remains a strong baseline,
particularly for unseen systems.

These studies primarily emphasize ranking preservation,
score estimation, or reducing the number of evaluations.
Our objective differs: each purchased evaluation incurs a
dollar cost, and the resulting observations must support
selection of an agent that satisfies a deployment-cost cap.
We measure whether the selected agent is truly affordable
and whether its accuracy difference from the best affordable
contender remains statistically unresolved, rather than
whether the full leaderboard ordering is reproduced.

\subsection{Budget Allocation and Configuration Search}

Budget allocation methods provide an alternative to selecting
individual informative tasks.
Successive halving repeatedly allocates evaluation resources
among candidates and eliminates weaker performers~\cite{sh},
building on sequential elimination methods~\cite{karnin}.
Hyperband extends this approach to configuration
search~\cite{hyperband}.
We adapt successive halving to agent evaluations with
unequal per-task dollar costs.

We also compare learned and heuristic acquisition strategies.
EZR uses lightweight active learning to identify promising
software configurations from limited measurements~\cite{ezrpaper,ezrcode}.
Our adaptation acquires complete agent runs, while a
cell-level Distance-to-Heaven (D2H)\cite{menzies2026moot} sampler prioritizes
individual agent--task evaluations using observed accuracy
and cost.
These approaches differ in their acquisition granularity
and use of observed information.

By evaluating these strategies under common dollar budgets,
we investigate whether more complex acquisition and stopping
rules improve affordable agent selection over simple allocation.
The policy adaptations and information constraints are
detailed in Section~\ref{shalf}.

\section{Data and Study Design}

\subsection{Trace Reconstruction and Inclusion Criteria}
\label{sec:data}

We reconstruct 58,197 run--task records from 379 of the 380
public HAL trace archives, spanning 11 benchmarks.
Each \emph{run} represents an executed agent configuration,
including its model, scaffold, and parameter settings.
We organize each benchmark into a matrix where rows represent
runs, columns represent tasks, and each \emph{cell} $(i,j)$
contains the outcome and execution cost of run $i$ on task $j$.
Repeated executions of the same configuration remain separate
rows to preserve run-to-run variation.

We exclude four benchmarks with incomplete outcome or token
representations.
For the remaining seven, we retain tasks shared across runs
and remove runs whose recorded outcomes cover fewer than half
their traced tasks or whose reconstructed costs are nonpositive.
Missing or nonpositive individual cell costs are imputed using
the corresponding run's median positive cell cost.
The resulting complete matrices contain 20,633 cells across
262 runs (Table~\ref{tab:resolution}).

Our primary software engineering benchmark, SWE-bench Verified
Mini, retains 34 runs evaluated on 48 of its original 50 tasks.
Recomputing accuracy on these shared tasks changes individual
scores by at most 2.33 percentage points without reversing
any strict pairwise ranking.
For other benchmarks, binary outcomes sometimes simplify
native grading: AssistantBench uses exact-match outcomes,
while CORE-Bench and SciCode use full task-level success
rather than finer subtask scores.

We distinguish three benchmark categories.
SWE-bench is our primary repository-maintenance
benchmark, evaluating agents that repair issues
in existing software repositories.
CORE-Bench, SciCode, and USACO provide
programming-oriented comparisons, while GAIA,
TAU-bench Airline, and AssistantBench serve as
general tool-use stress tests.
The additional benchmarks examine whether the
statistical and budget-allocation findings
extend beyond repository maintenance.

\subsection{Cost Reconstruction}
\label{sec:cost}

We reconstruct API execution costs from HAL token logs using
a standardized price sheet.
For each model call, let $I$, $R$, $W$, and $O$ denote inclusive
input, cache-read, cache-write, and output token counts,
with corresponding per-token prices $p_I,p_R,p_W,p_O$.
The reconstructed cost is

\begin{equation}
C=(I-R-W)p_I+Rp_R+Wp_W+Op_O.
\label{eq:cost}
\end{equation}

We normalize provider-specific token fields before applying
Equation~(\ref{eq:cost}).
When cache usage is available only at the run level, we
allocate cached tokens across tasks proportionally to their
prompt-token counts.

Because provider prices change over time, published
leaderboard costs may reflect different historical rate cards.
We recover token rates from single-model runs where possible
and standardize each model to its latest fitted pricing period,
using assumptions, documented in our artifact, when exact recovery is not possible.
This enables comparisons on a common price basis rather than
reproducing historical invoices.

Our reconstructed costs represent API token spend only;
they exclude hosting, external tools, and human-review costs.
Incomplete cache records and multi-model scaffolds introduce
additional uncertainty into task-level cost attribution.

\subsection{Paired Ranking Analysis}
\label{sec:paired}

Let $Y_{ij}\in\{0,1\}$ indicate whether run $i$ succeeds
on task $j$.
We compare run accuracy using the exact two-sided
McNemar test~\cite{mcnemar}, which accounts for paired
outcomes on the same tasks.
We apply Benjamini--Hochberg (BH) correction~\cite{bh}
at false discovery rate $q=0.05$ separately within
each benchmark, covering 5,371 pairwise comparisons.

We call a pair \emph{unresolved} when its corrected test
does not reject equal success probabilities.
This indicates insufficient evidence to distinguish the
runs, not proof of equivalent performance.

We also construct each benchmark's accuracy--cost Pareto
frontier, retaining runs for which no cheaper run achieves
equal or higher observed accuracy.
A \emph{frontier step} connects adjacent runs ordered by
increasing cost; it is unresolved when their paired
accuracy difference is not statistically significant.

We assess robustness using alternative tests, bootstrap
confidence intervals, and Bonferroni correction\cite{bonferroni1936teoria}.
We additionally examine whether bootstrap intervals fall
within specified accuracy margins to distinguish statistical
non-rejection from evidence of practical equivalence~\cite{phillips1990power, lakens2017equivalence}.

\begin{table}[!t]
\caption{Retained matrices and ranking resolution,
grouped by benchmark domain.
$k$: run rows; $n$: tasks;
Unres.: unresolved pairs after BH correction;
Steps: unresolved/all accuracy--cost frontier steps;
Floor: best-case exact-test gap
($100(6/n)$ pp, uncorrected;
Section~\ref{sec:rq1}).}
\label{tab:resolution}
\centering\small\setlength{\tabcolsep}{3.5pt}
\input{tables/resolution}
\end{table}

\subsection{Evaluation Budget and Deployment Cap}
\label{sec:budget}

We simulate budget-constrained evaluation using the retained
historical outcome and cost matrices.
Purchasing cell $(i,j)$ reveals its outcome $Y_{ij}$ and
incurs its reconstructed cost $C_{ij}$.
Let $E=\sum_{i,j}C_{ij}$ denote the cost of exhaustively
evaluating every run on every task.
We test evaluation budgets of 1\%, 2.5\%, 5\%, 10\%,
and 25\% of $E$, rounded to the nearest dollar.

We distinguish two constraints:
the \emph{evaluation budget} limits spending on
comparing candidate agents, while the
\emph{deployment cap} $d$ limits the reconstructed
API cost of executing the selected agent across
all retained benchmark tasks.
We use this complete-benchmark cost as a
workload-level proxy for agent operating expense.

For each benchmark, we consider three deployment
caps at the 25th, 50th, and 100th percentiles
of true reconstructed complete-run costs.
The 100th-percentile cap admits all retained runs.

After evaluation, each policy recommends the run with
the highest estimated accuracy whose estimated
complete-run cost does not exceed $d$.
Unless otherwise specified, accuracy is estimated from
observed task outcomes, and complete-run cost is estimated
as $n$ times the observed mean cell cost.
Runs without observations receive an accuracy prior of
0.5 and a pooled mean cost estimate from observed cells.

\subsection{Selection Quality and Uncertainty}
\label{sec:score}

Our objective is to select an affordable agent whose
accuracy is not statistically distinguishable from the
best affordable contender.

For deployment cap $d$, let $b_d$ denote the most accurate
run whose actual complete-run cost does not exceed $d$.
Let $\mathcal{U}_d$ contain all truly affordable runs
whose BH-corrected accuracy differences from $b_d$
are unresolved, including $b_d$ itself.

If a policy selects run $\hat{i}_d$, its
\emph{selection score} is

\begin{equation}
Q=\frac{1}{|\mathcal{D}_{\rm valid}|}
\sum_{d\in\mathcal{D}_{\rm valid}}
\mathbf{1}[\hat{i}_d\in\mathcal{U}_d],
\label{eq:quality}
\end{equation}

where $\mathcal{D}_{\rm valid}$ contains deployment caps
for which the policy makes a recommendation.
Caps without a recommendation are omitted, and $Q$
is missing if no recommendation is made.

Because $Q$ conditions on a recommendation, we also
report an \emph{all-caps score} that counts omitted
recommendations as failures.
To assess the quality of selected agents beyond statistical
non-rejection, we additionally measure true affordability,
accuracy regret relative to $b_d$, and an
\emph{exact-best score} that requires selecting a truly
affordable run attaining the highest observed accuracy.

These complementary metrics are important because a high
$Q$ on a low-resolution benchmark does not establish
that the selected agent matches the best agent on future tasks.

Our primary experiments use 300 randomized replicates
per condition.
We compare policy differences using paired bootstrap
intervals.
To estimate the spending required to reach selection
quality $Q=0.80$, we additionally run 200 replicates
across ten budget levels from 0.5\% to 40\% of $E$
and interpolate the first crossing of each policy's
mean quality curve.
Additional thresholds and sensitivity analyses are
reported in the replication artifact.

\subsection{Evaluation Policies and Information Boundaries}
\label{shalf}

We compare simple allocation baselines, adaptive
selection strategies, and cost-aware extensions.

\noindent
\textbf{Simple allocation.}
Our non-adaptive baselines include random-cell sampling,
random-task sampling (evaluating all runs on sampled tasks),
round robin across runs, and cheapest-first acquisition.

We also adapt \emph{successive halving} (SH)~\cite{sh},
which assigns equal dollar-spending
targets to surviving runs within each round.
Each run purchases unseen tasks until its spending
reaches its target, with individual purchases
allowed to exceed the local target if sufficient
global evaluation budget remains.
Unused funds remain available globally rather
than carrying over as run-specific allocations.

After each round, SH eliminates the lower-scoring
half of surviving runs based on observed accuracy.
Runs without observations rank below observed runs,
with ties preserving the current survivor order.
Consequently, at tight budgets, some candidates
may be eliminated without an accuracy observation.

\smallskip
\noindent
\textbf{Selective task sampling.}
We adapt mid-range filtering (MR)~\cite{ndzomga},
which initially spends 10\% of its evaluation budget on
random-cell probes to estimate task success rates.
Subsequent purchases prioritize tasks with estimated
success rates between 0.30 and 0.70, using remaining
tasks when necessary.
An oracle variant uses true full-matrix task success
rates as a diagnostic upper-information condition.

\smallskip
\noindent

\textbf{Learned and heuristic acquisition.}
Our row-level EZR adaptation~\cite{ezrpaper,ezrcode}
uses a best-versus-rest~\cite{tiao2021bore} Naive Bayes classifier\cite{webb2010naive}
over model identifier, model family, provider,
reasoning effort, scaffold, and token price rates.
It begins with up to four complete runs and
subsequently acquires one unobserved run at a time.

Cell-level Distance-to-Heaven (D2H) instead
prioritizes individual agent--task evaluations.
It first attempts two randomly selected task
purchases per run, subject to budget feasibility.
For each run $i$, D2H computes

\begin{equation}
d_i = \sqrt{\frac{(1-\tilde{a}_i)^2+
                    \tilde{c}_i^2}{2}},
\qquad
s_i = \frac{w_i}{\sqrt{n_i+1}},
\label{eq:d2h}
\end{equation}

where $\tilde{a}_i$ and $\tilde{c}_i$ are
min--max normalized observed accuracy and
$\log(1+\hat{c}_i)$ across candidate runs.
Here, $\hat{c}_i$ estimates complete-run cost
as the observed mean cell cost multiplied by
the benchmark task count, and $n_i$ is the
number of purchased cells.

The weight $w_i$ is 1 for the
$\max(1,\lfloor\sqrt{k}\rfloor)$ runs with
smallest $d_i$, and 0.15 otherwise,
where $k$ is the number of candidate runs.
D2H visits runs in descending acquisition score
$s_i$, breaking ties randomly, and selects an
unseen task uniformly at random.
Under advance cost checks, unaffordable purchases
are skipped in favor of the next candidate.

We also test additional acquisition heuristics,
cost estimators, and early-stopping rules.
Detailed variants and ablations are provided
in the replication artifact.

\smallskip
\noindent
\textbf{Cost visibility.}
Our primary replay assumes \emph{advance cost checks}:
policies can inspect the exact historical cost $C_{ij}$
before purchasing a cell and skip purchases exceeding
the remaining budget.
This is an idealized setting because an agent's actual
token usage is generally unknown before execution.

We therefore also evaluate a prospective setting
without advance cost checks.
Here, policies observe execution costs only after
purchasing cells and may exceed their nominal budgets.
All oracle conditions are separated from deployable
policies that rely only on purchased observations.

\section{RQ1: Statistical Support for Accuracy Gains}
\label{sec:rq1}

\subsection{Resolution of Accuracy-Cost Trade-offs}

Paying more for a higher-ranked agent does not necessarily
yield a statistically distinguishable accuracy improvement.
On SWE-bench Verified Mini, three of four steps along the
accuracy-cost Pareto frontier remain statistically unresolved.
Among 34 runs evaluated on 48 shared tasks, 225 of 561
pairwise comparisons (40.1\%) are unresolved after
BH correction.

Figure~\ref{fig:swe-tradeoffs} illustrates this trade-off.
The cheapest contender statistically unresolved from the
leader costs \$34.16, compared with \$170.78 for the leader,
despite trailing by 14.6 percentage points in accuracy.
The available evidence cannot establish whether this
fivefold cost premium is justified.
This does not imply equivalent performance, but identifies
a cheaper candidate worth further evaluation.

The limitation extends beyond software maintenance.
Across seven HAL benchmarks, 27 of 35 accuracy-cost
frontier steps remain unresolved (Table~\ref{tab:resolution}).
Six benchmarks contain at least one unresolved step,
with USACO the only exception.
Across all 5,371 pairwise comparisons, unresolved fractions
range from 21.9\% on USACO to 100\% on SciCode and
AssistantBench.
Thus, leaderboard rankings often imply finer distinctions
than the available statistical evidence supports.

\subsection{Why Small Benchmarks Cannot Resolve Accuracy Gaps}

Limited statistical resolution is partly a mathematical
consequence of benchmark size~\cite{card2020little}.
Consider two runs evaluated on $n$ shared tasks.
If they disagree on $m$ tasks and every disagreement favors
the same run, the exact two-sided McNemar test yields
$p=2(1/2)^m$.
At significance level $\alpha=0.05$, at least six such
disagreements are needed to reject equal success probabilities.
Therefore, the smallest observable accuracy gap that an
uncorrected exact test can resolve is $6/n$~\cite{connor1987sample}.
Multiple-testing correction and disagreements in both
directions can only increase this requirement.

For SWE-bench ($n=48$), this best-case resolution floor is
12.5 percentage points.
In practice, competing agents often succeed on different
subsets of tasks, further reducing statistical power.
Simulations incorporating observed disagreement patterns
estimate that SWE-bench requires a true accuracy gap of
20.7 percentage points to achieve 80\% rejection power
at uncorrected $\alpha=0.05$
(Table~\ref{tab:robust}).

The same limitation is more severe on smaller or
lower-variance benchmarks.
AssistantBench ($n=33$) has an exact-test floor of
18.2 percentage points, exceeding its entire observed
accuracy range of 15.2 points.
Consequently, no pair can pass even an uncorrected exact
test under the retained binary outcomes.
On SciCode ($n=65$), the 9.2-point resolution floor matches
the observed score range, and no comparisons survive
BH correction.

\subsection{Robustness and Practical Interpretation}

\begin{table}[!b]
\caption{Ranking-resolution sensitivity checks
grouped by benchmark domain.
Exact and Boot.: unresolved percentages without
correction under exact McNemar and 95\% bootstrap
intervals.
Bonf.: Bonferroni-corrected unresolved percentages.
Within: percentage of 90\% intervals inside
$\pm10$ percentage points.
MDE: simulated 80\%-power accuracy gap in
percentage points.}
\label{tab:robust}
\centering\small\setlength{\tabcolsep}{3pt}
\input{tables/robustness}
\end{table}

Low ranking resolution is not solely an artifact of
BH correction.
Without multiple-testing correction, 21.0\% to 100\%
of run pairs remain unresolved under exact McNemar tests,
with similar results from unadjusted bootstrap intervals
(Table~\ref{tab:robust}).
Alternative tests and Bonferroni correction provide
additional sensitivity checks.

Importantly, statistical non-rejection does not establish
practical equivalence.
On SWE-bench, only 4.8\% of run pairs have 90\% bootstrap
accuracy-difference intervals entirely within
$\pm10$ percentage points.
Thus, many unresolved comparisons still permit
differences large enough to matter in deployment.

\smallskip
\noindent

\begin{rqtakeaway}
\textbf{RQ1 Takeaway.}
Most accuracy--cost frontier improvements remain statistically
unresolved due to limited benchmark size and disagreements
between agents.
Preserving leaderboard order alone cannot justify paying
more for a higher-ranked agent; selection must account
for accuracy uncertainty and cost.
\end{rqtakeaway}

\section{RQ2: Reliability of Agent Cost Estimates}
\label{sec:rq2}

Reliable agent selection requires comparable execution costs
and accurate estimates of full-benchmark deployment spend.
We examine two sources of uncertainty: discrepancies in
published leaderboard costs and bias introduced when
deployment costs are estimated from sampled tasks.

\subsection{Discrepancies in Reported Execution Costs}

Across 333 runs with positive published and reconstructed
costs, the published-to-reconstructed cost ratio has
a median of 1.00, a 95th percentile of 5.00,
and a maximum of 11.7$\times$.
These discrepancies reflect accounting inconsistencies
and historical pricing differences.

\textbf{Prompt-cache accounting.}
In two Anthropic agent runs, logged token totals are
consistent with cached tokens being counted twice,
producing recorded-to-reconstructed cost ratios of
11.7$\times$ and 9.9$\times$. The inset of Figure~\ref{fig:swe-tradeoffs} illustrates the 11.7$\times$ cache-accounting discrepancy.
For one run, HAL reports \$1,789.67, compared with
\$180.13 under our standardized reconstruction.
An execution of the same configuration approximately
five hours earlier has substantially different reported
costs despite similar reconstructed spend.
This illustrates how inconsistent cache accounting
can distort comparisons between agent runs.

\textbf{Historical pricing.}
Leaderboard costs also reflect different provider price
periods.
For 16 runs using OpenAI's \texttt{o3}, recorded costs
reflect an earlier \$10/\$40 per-million-token
input/output price sheet, compared with the \$2/\$8
rates used in our standardized reconstruction.
This produces a 5.0$\times$ difference attributable
to pricing rather than token consumption.
Some runs also applied legacy rates after the price
change, further complicating historical comparisons.

\textbf{Reconstruction limits.}
Cost reconstruction is constrained by incomplete traces
and multi-model scaffolds.
Forty-two extracted runs lack recoverable token records,
while multi-model executions do not always preserve
sufficient information for precise cache attribution.
Regression checks recover exact input/output rates for
14 model identifiers, but cache-rate assumptions remain
necessary where the available logs are insufficient.
Thus, reconstructed costs provide a consistent comparison
basis rather than exact historical invoices.

\subsection{Sampling Bias in Deployment-Cost Estimates}

Even with standardized prices, evaluating only a subset
of tasks can bias estimates of complete-run cost.
We compare the estimated-to-true cost ratio across
sampling strategies at a 10\% evaluation budget
(Table~\ref{tab:bias}).

\begin{table}[t]
\caption{Median estimated/true complete-run cost ratio
at a 10\% evaluation budget (100 replicates).
Values below 1 indicate underestimation.
Oracle conditions use full-matrix task pass rates.}
\label{tab:bias}
\centering\small\setlength{\tabcolsep}{3pt}
\input{tables/bias}
\end{table}

Under oracle mid-range filtering, which selects tasks
with full-matrix pass rates between 0.30 and 0.70,
the median estimated-to-true cost ratio falls to
0.796 on SWE-bench and 0.834 on CORE-Bench.
These correspond to underestimates of 20.4\%
and 16.6\%, respectively.
By contrast, random-task sampling yields ratios of
0.985 and 0.933 on the same benchmarks, despite
purchasing fewer cells.

Tasks selected for accuracy discrimination may not
represent the full distribution of execution costs.
Across 34 retained SWE-bench runs, the 20 tasks with
pass rates below 0.30 incur a mean cost of \$4.38
per run--task cell, compared with \$2.59 for the
27 mid-range tasks (pass rates 0.30--0.70).
Thus, the hardest issues cost 1.69$\times$ more per attempt and fail
more often.
For a maintenance team, the cost of failed attempts falls on the
hardest items in the backlog.
A sampler that skips these issues to sharpen accuracy comparisons also
skips the most expensive part of the workload.

The practical mid-range sampler, which includes an
initial random probe and purchases tasks outside
the preferred difficulty band when necessary,
shows substantially less bias on SWE-bench (0.973).
Separating cost estimation from accuracy-driven sampling
also improves cost ratios: a dedicated uniform probe
yields 1.006 on SWE-bench and 1.037 on CORE-Bench,
although residual bias remains on SciCode (0.848).
These results suggest that representative cost
observations are important when task selection is
optimized for accuracy discrimination.

\smallskip
\noindent

\begin{rqtakeaway}
\textbf{RQ2 Takeaway.}
Leaderboard costs can be wrong by up to 11.7$\times$ due to cache
accounting and old price sheets.
Hard issues cost more and fail more, so samplers that skip them
underestimate deployment cost by up to 20\%.
Teams need standardized costs and a separate, uniform cost sample.
\end{rqtakeaway}

\section{RQ3: Budgeted Agent Selection}
\label{sec:rq3}

\begin{figure*}[t]
\centering
\includegraphics[width=.8\textwidth]{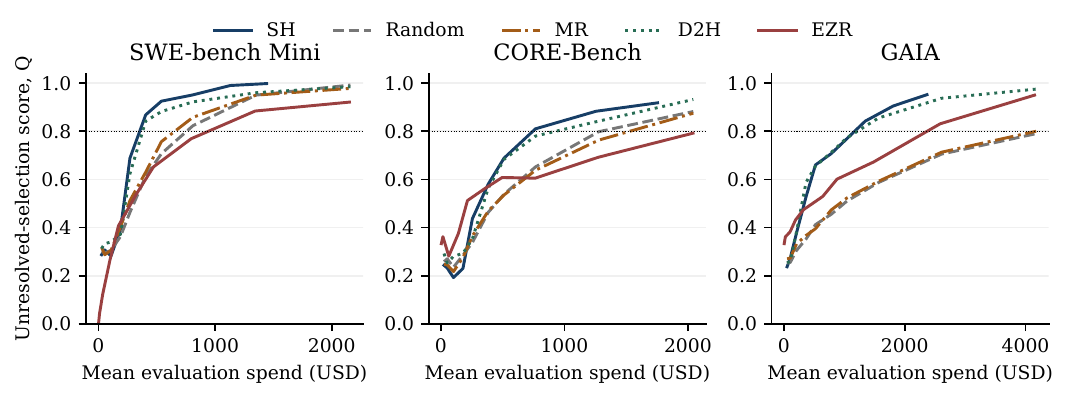}
\caption{Mean selection score $Q$ versus reconstructed evaluation
spend on three benchmarks (200 replay replicates per point).
The dotted line marks the target $Q=0.80$.}
\label{fig:curves}
\end{figure*}

\begin{table*}[t]
\caption{Interpolated mean evaluation spend (USD)
required to reach $Q=0.80$ under advance cost checks,
grouped by benchmark domain.
Bold indicates the lowest spend among five policies.
A dash indicates the target was not reached within
40\% of exhaustive evaluation cost.}
\label{tab:cost}
\centering\small\setlength{\tabcolsep}{10pt}\input{tables/cost}
\end{table*}

\subsection{Evaluation Cost of Competitive Selection}

Successive halving (SH) provides a strong baseline for
selecting affordable, competitive agents under limited
evaluation budgets.
When policies can check individual task costs before execution,
SH reaches the uncertainty-aware selection targets
$Q=0.80$ and $Q=0.90$ on all seven benchmarks.
For $Q=0.80$, it achieves the lowest interpolated
evaluation spend on six benchmarks and ties on the
seventh (Table~\ref{tab:cost}).

On SWE-bench Verified Mini, exhaustively evaluating
34 runs across 48 tasks costs \$5,399.89.
SH reaches $Q=0.80$ after spending \$353.62,
just 6.55\% of exhaustive evaluation cost.
Cell-level D2H follows closely at \$379.19,
whereas row-level EZR requires \$942.20.
These differences describe crossings of interpolated
mean curves rather than statistically established
differences in minimum evaluation cost.

The result extends across the programming-oriented
benchmarks.
SH achieves the lowest interpolated spend on
SWE-bench, USACO, and CORE-Bench, and ties with
other policies on SciCode.
On CORE-Bench, SH requires \$744.44 compared with
\$1,292.27 for random-task sampling.
Its advantage is smaller on USACO, where the two
policies require nearly identical budgets
(\$72.69 and \$72.93).
Among general tool-use benchmarks, SH reaches
$Q=0.80$ on GAIA for \$1,171.54, while random-task
sampling does not reach the target within 40\%
of exhaustive evaluation spend
(Figure~\ref{fig:curves}).

However, selection quality also depends on
benchmark resolution.
On SciCode, all corrected pairwise accuracy
comparisons are unresolved, so any truly
affordable selection receives full credit under $Q$.
Several policies consequently reach the target
at the first tested budget level.
Their apparent tie therefore reflects the
tested budget grid and limited statistical
resolution, not necessarily identical minimum
evaluation costs.
This illustrates why selection scores must be
interpreted alongside ranking uncertainty
(Section~\ref{sec:rq1}).

\subsection{Selection Quality Beyond Statistical Ties}

\begin{table}[t]
\caption{Successive halving selection quality at a 25\%
evaluation budget. Aff.: percentage of selections
that are truly affordable; Regret: mean accuracy
regret in percentage points among affordable
selections; Exact best: percentage selecting a
truly affordable run with the highest observed accuracy.}
\label{tab:selection-sensitivity}
\centering\small
\setlength{\tabcolsep}{4pt}
\begin{tabular}{lrrr}
\toprule
Benchmark & Aff. (\%) & Regret (pp) & Exact best (\%) \\
\midrule
SWE-bench Mini & 100.0 & 1.4 & 60.5 \\
USACO & 99.8 & 0.5 & 91.3 \\
SciCode & 100.0 & 0.0 & 99.7 \\
CORE-Bench & 93.8 & 3.2 & 58.3 \\
\midrule
GAIA & 91.2 & 0.8 & 79.3 \\
TAU Airline & 99.8 & 2.5 & 58.0 \\
AssistantBench & 100.0 & 0.8 & 74.8 \\
\bottomrule
\end{tabular}
\end{table}

Our primary selection score $Q$ becomes less
discriminative when many accuracy comparisons
are statistically unresolved.
We therefore assess whether high $Q$ corresponds
to genuinely affordable selections with small
observed accuracy regret (Table~\ref{tab:selection-sensitivity}).

At a 25\% evaluation budget, SH selects truly
affordable runs in 91.2\%-100\% of cap decisions
across benchmarks, with mean accuracy regret
between 0.0 and 3.2 percentage points.
On SWE-bench, all selections are affordable,
with mean regret of 1.4 percentage points,
although only 60.5\% select the exact
highest-scoring affordable run.

These results indicate that SH often identifies
affordable candidates with low observed regret,
even when it does not identify the exact-best
configuration.

\subsection{Evaluation Without Advance Cost Checks}

The preceding results assume advance knowledge of
each cell's execution cost.
In live evaluations, however, token consumption is
generally unknown until an agent completes its task.
We therefore repeat the replay without advance cost
checks, allowing purchases to exceed the nominal
evaluation budget.

SH remains competitive but no longer leads on
nearly every benchmark.
Under the all-caps $Q=0.80$ criterion, it achieves
the lowest mean actual spend on four of seven
benchmarks.
Random-task sampling leads on USACO, mid-range
filtering on SciCode, and D2H on GAIA.

Unknown execution costs also introduce budget overruns.
At a 10\% SWE-bench evaluation budget, every SH
replicate exceeds its allocation, although the
mean overrun is only \$3.04.
Row-level EZR incurs larger overruns because acquiring
a complete run commits the evaluator to many task
executions at once.
Cell-level policies offer finer budget control by
purchasing individual agent--task evaluations.

These results indicate that both cost visibility
and acquisition granularity influence the effectiveness
of budgeted agent selection.

\subsection{Do More Complex Acquisition Heuristics Help?}

Variants of row-level EZR yield mixed results
across benchmarks.
Although expected-improvement (EI)~\cite{jones1998efficient}
acquisition improves selection quality in several settings,
it reduces SWE-bench $Q$ by 0.132 at the 25\% budget
(95\% CI $[-0.169,-0.097]$).
Alternative token-price feature variants likewise
provide no consistent benefit.

Additional heuristics also fail to consistently improve
cell-level D2H.
Adding cost-normalized acquisition weights and mixed-outcome
task prioritization reduces $Q$ at a 25\% budget on
SWE-bench ($0.925$ versus $0.968$ for D2H),
USACO ($0.841$ versus $0.964$), and
CORE-Bench ($0.769$ versus $0.838$).
Thus, simpler acquisition rules can outperform
more elaborate alternatives.

Cost-estimation choices also affect selection.
Using median rather than mean cell costs can
underestimate complete-run costs for right-skewed
execution distributions.
Under median-based estimation, D2H reaches
$Q=0.80$ at lower spend than SH on SWE-bench
(\$449.04 versus \$477.87), while no policy
reaches the target on GAIA within the tested range.
By contrast, stress tests of token rates and
cache allocation leave SH's leading position
largely unchanged in the examined cases.

\subsection{Early Stopping and Selection Quality}

Finally, we examine whether stopping evaluation
when predicted improvements become small can
reduce spending without compromising selection.

Our stopping rule halts evaluation when every
candidate's predicted one-step accuracy improvement
falls below $\lambda(6/n)$, where $6/n$ is the
benchmark's uncorrected exact-test resolution floor.
We vary the threshold scale $\lambda$ to examine
the trade-off between evaluation spending and
selection quality.

Early stopping produces a clear cost--quality
trade-off.
At the 25\% SWE-bench budget, stopping at a
threshold scale of 0.25 reduces budget use by
16.1 percentage points but lowers $Q$ from
0.756 to 0.614.
On benchmarks with limited ranking resolution,
stopping can save substantially more budget with
little change in selection score.

However, this apparent efficiency can reflect
the inability of a benchmark to distinguish
candidate agents rather than successful
identification of the best candidate.
Aggressive stopping is therefore difficult to
justify when meaningful accuracy differences
remain unresolved.

\smallskip
\noindent
\begin{rqtakeaway}
\textbf{RQ3 Takeaway.}
Under advance cost checks, successive halving reaches
the uncertainty-aware SWE-bench selection target using
6.55\% of exhaustive evaluation spend.
Its advantage narrows under unknown execution costs
and stricter selection criteria, while more complex
heuristics offer inconsistent improvements.
\end{rqtakeaway}

\section{Threats to Validity}
\label{sec:threats}

\textbf{Outcome validity.}
We treat SWE-bench's test-based ``resolved'' label as ground truth.
Prior work shows that some test-passing patches are incorrect or only
plausible~\cite{wang2026solved, aleithan2024swebenchplus}.
Such errors add noise to accuracy estimates and can further reduce
ranking resolution.
Our cost findings do not depend on this label.

\textbf{Dataset validity.}
HAL provides a convenience sample of agent executions,
and our completeness filters exclude benchmarks and
runs with insufficient traces or outcomes.
SWE-bench analysis uses 48 shared tasks rather than
the full 50-task subset.
Other benchmarks sometimes use simplified binary
outcomes instead of native grading metrics.
These choices limit generalizability, particularly
beyond repository-level maintenance tasks.

\textbf{Statistical validity.}
Paired tests treat benchmark tasks as evaluation units,
potentially overlooking dependencies between issues
from the same repository~\cite{du2025dependeval, miller2024adding}.
Because comparisons share runs and tasks, BH correction
does not guarantee false discovery rate control under
arbitrary dependence~\cite{benjamini2001control}.
We therefore report alternative tests and correction
procedures as sensitivity checks.
Moreover, statistical non-rejection does not establish
practical equivalence.
Only four repeated-configuration groups are available,
limiting conclusions about agent execution variability~\cite{ouyang2025empirical}.

\textbf{Cost validity.}
Reconstructed costs rely on standardized pricing,
cache-allocation assumptions, and occasional imputation.
Incomplete traces and multi-model scaffolds further
limit exact attribution.
Our estimates capture API token spending but exclude
hosting, external tools, and human-review costs.
Consequently, reconstructed dollar amounts support
comparative analysis rather than exact historical
invoices or future deployment quotes.

\textbf{Replay and selection validity.}
Budget policies operate on fixed historical outcomes
rather than fresh agent executions.
The primary replay assumes advance knowledge of
individual task costs, which is generally unavailable
in live evaluations.
Our prospective replay relaxes this assumption but
can exceed nominal budgets.
Furthermore, the selection score $Q$ credits unresolved
accuracy differences and omits caps without a
recommendation, potentially favoring policies on
low-resolution benchmarks.
We address these limitations through affordability,
all-caps, regret, and exact-best analyses, but
prospective validation remains necessary.

\textbf{Comparison validity.}
Evaluated strategies differ in acquisition granularity,
initialization, features, and cost estimators.
Their observed performance therefore reflects complete
policy implementations rather than isolated algorithmic
components.
Because all strategies were explored on the same
historical corpus, our findings establish comparative
evidence within this setting, not universal superiority
of successive halving.

Our SH implementation also uses soft per-run
spending targets and does not guarantee an
initial observation for every candidate.
At tight budgets, runs receiving no observations
can be eliminated before their accuracy is measured,
potentially disadvantaging expensive configurations.

\section{Discussion and Practical Implications}
\label{sec:discussion}

Our findings suggest that evaluating software engineering
agents should focus on supporting reliable, affordable
selection decisions rather than reproducing leaderboard
rankings alone.
We discuss three implications for agent evaluation and
software maintenance practice.

\subsection{From Leaderboard Rankings to Agent Selection}

\textbf{Report supported accuracy--cost trade-offs.}
Agent leaderboards should report accuracy uncertainty
alongside execution costs, rather than relying solely
on point estimates and ranks.
As RQ1 demonstrates, higher-ranked agents frequently
lack statistically distinguishable gains over cheaper
alternatives.
For software teams selecting maintenance agents, an
observed accuracy improvement should therefore be
weighed against its additional execution cost and
the uncertainty of the measured difference.
Statistically unresolved comparisons do not establish
equivalent patch quality; instead, they identify
candidates that warrant further evaluation before
committing to a more expensive configuration.

\subsection{Reliable Cost Measurement and Estimation}

\textbf{Separate evaluation budgets from deployment costs.}
Agent evaluation involves deciding both how much to
spend comparing candidate systems and how much to
spend executing the selected agent.
These costs should be reported separately, with
task-level token usage, cache accounting, and
model pricing preserved for reproducibility.
Standardized price sheets can prevent historical
pricing differences from distorting comparisons.

\textbf{Validate costs independently of accuracy sampling.}
Tasks selected for their ability to distinguish agent
accuracy may not represent the execution costs of
the complete workload.
As RQ2 shows, difficulty-based filtering can
underestimate full-benchmark costs, while separate
uniform cost probes reduce this bias in several cases.
Evaluators should therefore check whether
accuracy-oriented task subsets also provide
representative cost estimates before using them
to enforce deployment budgets.

\subsection{Toward Budget-Aware Agent Evaluation}

\textbf{Establish simple adaptive allocation baselines.}
Successive halving provides a strong starting point
for evaluating multiple agent configurations under
limited budgets.
By allocating initial resources broadly and
progressively concentrating evaluation on promising
candidates, it can substantially reduce evaluation
spending without requiring complex acquisition models.
Our results suggest that additional cost-aware
heuristics and early-stopping rules should be
justified by consistent improvements in selection
quality over such simple baselines.

However, budget-aware evaluation must also reflect
the unpredictability of agent execution.
In practice, multi-turn tool-use trajectories have
unknown costs before completion, and successive
halving's advantage narrows when advance cost checks
are unavailable.
Practical evaluation systems should therefore track
realized spending, account for possible overruns,
and assess selection quality beyond statistical ties.

\textbf{Validate selection on live backlogs.}
A natural next step is to run budgeted selection on a team's own
issue backlog, with patch review as the outcome measure.
Our analysis considers complete agent configurations,
including models and scaffolds, but measures historical
benchmark outcomes and API token costs rather than
the downstream quality of deployed patches.
Future work should validate budgeted selection
through fresh repository-level executions that
account for agent stochasticity, patch correctness,
maintainability, and operational costs.
This would establish whether efficient benchmark
selection translates into reliable software
maintenance decisions.

\smallskip
\noindent
\textbf{Closing perspective.}
Across seven HAL benchmarks, 27 of 35 accuracy-cost
frontier steps remain statistically unresolved,
while successive halving reaches our uncertainty-aware
selection target on SWE-bench using only 6.6\%
of exhaustive evaluation spend under advance cost checks.
Together with the cost-accounting and sampling-bias
findings, these results motivate a decision-oriented
approach to agentic software engineering evaluation.
Progress should be measured not only by higher
benchmark scores, but by how reliably and economically
those scores support agent-selection decisions.

\section*{Data Availability}
The source execution traces are publicly available in the HAL dataset~\cite{hal}:
\url{https://huggingface.co/datasets/agent-evals/hal_traces}.
The study uses the released task outcomes and token-usage records to construct its analysis matrices.

\bibliographystyle{IEEEtran}
\bibliography{references}
\end{document}

%% file: tables/resolution.tex
\begin{tabular}{lrrrrr}
\toprule
Benchmark & $k$ & $n$ & Unres. (\%) & Steps & Floor (pp) \\
\midrule
\multicolumn{6}{@{}l}{\emph{Repository maintenance}} \\
SWE-bench Mini & 34 & 48 & 40.1 & 3/4 & 12.5 \\
\midrule
\multicolumn{6}{@{}l}{\emph{Programming-oriented tasks}} \\
USACO & 15 & 307 & 21.9 & 0/4 & 2.0 \\
SciCode & 38 & 65 & 100.0 & 2/2 & 9.2 \\
CORE-Bench & 61 & 45 & 67.5 & 7/8 & 13.3 \\
\midrule
\multicolumn{6}{@{}l}{\emph{General tool-use}} \\
GAIA & 35 & 165 & 32.4 & 7/9 & 3.6 \\
TAU Airline & 47 & 50 & 79.8 & 6/6 & 12.0 \\
AssistantBench & 32 & 33 & 100.0 & 2/2 & 18.2 \\
\bottomrule
\end{tabular}

%% file: tables/robustness.tex
\begin{tabular}{lrrrrr}
\toprule
Benchmark & Exact & Boot. & Bonf. & Within & MDE \\
\midrule
\multicolumn{6}{@{}l}{\emph{Repository maintenance}} \\
SWE-bench Mini & 36.9 & 33.2 & 62.6 & 4.8 & 20.7 \\
\midrule
\multicolumn{6}{@{}l}{\emph{Programming-oriented tasks}} \\
USACO & 21.0 & 20.0 & 36.2 & 21.9 & 6.6 \\
SciCode & 97.7 & 83.4 & 100.0 & 75.7 & 12.2 \\
CORE-Bench & 58.7 & 52.9 & 95.4 & 2.1 & 24.9 \\
\midrule
\multicolumn{6}{@{}l}{\emph{General tool-use}} \\
GAIA & 31.1 & 29.4 & 52.1 & 10.9 & 13.3 \\
TAU Airline & 68.1 & 64.0 & 96.4 & 0.2 & 30.2 \\
AssistantBench & 100.0 & 94.6 & 100.0 & 47.8 & 22.8 \\
\bottomrule
\end{tabular}

%% file: tables/bias.tex
\begin{tabular}{lrrrr}
\toprule
Estimator / sampler & SWE & USACO & SciCode & CORE \\
\midrule
Oracle MR & 0.796 & 0.986 & 0.822 & 0.834 \\
Probed MR & 0.973 & 0.994 & 0.991 & 0.934 \\
Random tasks & 0.985 & 0.992 & 0.909 & 0.933 \\
Oracle MR + probe & 0.793 & 0.988 & 0.827 & 0.827 \\
Oracle MR + cost model & 0.795 & 0.965 & 0.714 & 0.805 \\
Oracle MR + both & 0.982 & 0.981 & 0.884 & 1.055 \\
Oracle MR, separate costs & 1.006 & 1.007 & 0.848 & 1.037 \\
\bottomrule
\end{tabular}

%% file: tables/cost.tex
\begin{tabular}{lrrrrr}
\toprule
Benchmark & SH & Random & MR & D2H & EZR \\
\midrule
\multicolumn{6}{@{}l}{\emph{Repository maintenance}} \\
SWE-bench Mini & \textbf{353.62} & 755.98 & 655.06
               & 379.19 & 942.20 \\
\midrule
\multicolumn{6}{@{}l}{\emph{Programming-oriented tasks}} \\
USACO & \textbf{72.69} & 72.93 & 83.12
      & 173.55 & -- \\
SciCode & \textbf{25.29} & \textbf{25.29}
        & \textbf{25.29} & \textbf{25.29} & 36.73 \\
CORE-Bench & \textbf{744.44} & 1,292.27 & 1,528.52
           & 931.82 & -- \\
\midrule
\multicolumn{6}{@{}l}{\emph{General tool-use}} \\
GAIA & \textbf{1,171.54} & -- & 4,174.73
     & 1,214.07 & 2,371.57 \\
TAU Airline & \textbf{156.11} & 235.16 & 218.63
            & 160.55 & 280.07 \\
AssistantBench & \textbf{81.87} & 200.61 & 158.50
               & 140.32 & 124.48 \\
\bottomrule
\end{tabular}